\documentclass[twocolumn,aps,pra]{revtex4}
\usepackage[T1]{fontenc}
\usepackage[latin1]{inputenc}
\usepackage{amsmath}
\usepackage{amssymb}
\usepackage{graphicx}

\makeatletter
\@ifundefined{textcolor}{}
{%
 \definecolor{BLACK}{gray}{0}
 \definecolor{WHITE}{gray}{1}
 \definecolor{RED}{rgb}{1,0,0}
 \definecolor{GREEN}{rgb}{0,1,0}
 \definecolor{BLUE}{rgb}{0,0,1}
 \definecolor{CYAN}{cmyk}{1,0,0,0}
 \definecolor{MAGENTA}{cmyk}{0,1,0,0}
 \definecolor{YELLOW}{cmyk}{0,0,1,0}
}

\usepackage{braket}
\usepackage{float}
\usepackage{tabularx}
\usepackage{booktabs}
\usepackage{amsfonts}\usepackage{epsf}
\usepackage{xcolor}
\usepackage{ifpdf}
\usepackage{epstopdf}
\usepackage{bm}

\makeatother

\begin{document}

\title{Theory of strongly paired fermions with arbitrary short-range interactions }

\author{Jianshen Hu$^{1}$}

\author{Fan Wu$^{1}$}

\author{Lianyi He$^{1}$}

\author{Xia-Ji Liu$^{2}$}

\author{Hui Hu$^{2}$}

\affiliation{$^{1}$ Department of Physics and State Key Laboratory of Low-Dimensional
Quantum Physics, Tsinghua University, Beijing 100084, China}

\affiliation{ $^{2}$ Centre for Quantum and Optical Science, Swinburne University
of Technology, Melbourne 3122, Australia}

\date{\today}
\begin{abstract}
We develop an effective field theory to describe the superfluid
pairing in strongly interacting fermions with arbitrary short-range
attractions, by extending Kaplan's idea of coupling fermions to a
fictitious boson state in Nucl. Phys. B \textbf{494}, 471 (1997).
This boson field is assigned with an unconventional kinetic term to recover
the exact scattering phase shift obtained either from scattering data
or model calculations. The theory works even if the explicit form of the interaction potential has not been constructed from scattering data. 
The contact boson-fermion coupling allows us
to go beyond mean-field to include Gaussian pair fluctuations, yielding
reliable predictions on equations of state. As an application, we
use our theory to address the non-univerisal ground-state energy of
strongly paired fermions, due to the non-trivial momentum dependence
of the phase shift characterized, for example, by effective range.
We find a good agreement between our predictions and recent quantum
Monte Carlo simulations on the effective-range dependence in both
three and two spatial dimensions. We propose that in cold-atom experiments,
the non-universal dependence in thermodynamics can be probed using
dark-state optical control of Feshbach resonances.
\end{abstract}
\maketitle

\section{Introduction}\label{s1}

In quantum many-body Fermi systems,
attractive inter-particle interaction leads to Cooper pairing and
superfluidity \cite{Review01,Review02,Review03,Exciton}. Increasing
the attraction strength will induce a crossover from a Bardeen-Cooper-Schrieffer
(BCS) superfluid state with largely overlapping Cooper pairs to a
Bose-Einstein condensate (BEC) of tightly bound molecules~\cite{Eagles,Leggett,NSR,BCSBEC1,BCSBEC2,BCSBEC3,BCSBEC4}.
Ultracold atomic Fermi gases near magnetic-field-tuned Feshbach resonances
provide clean systems to demonstrate the BCS-BEC crossover \cite{Regal2004,Zwierlein2004,Kinast2004}
and explore novel many-body phenomena such as universal thermodynamics
\cite{Ho2004,Hu2007,Thomas2009,Nascimbene2010,Navon2010,Horikoshi2010,Ku2012}.

The universal properties of an ultracold Fermi gas stem from its simple
form of the scattering phase shift in the dilute limit, characterized
by a large $s$-wave scattering length and negligible effective range~\cite{TH01,TH02,TH03,TH04,TH05,TH06,TH07,TH08,TH09,TH10,TH11,TH12,EOSmc1,EOSmc2,EOSmc3,EOSmc4}.
In realistic systems, however, non-universal effects could be important
due to the non-trivial momentum dependence in the phase shift, which
leads to, for example, nonzero effective range. For instance, neutron
superfluid with a large $s$-wave scattering length $a_{{\rm nn}}\simeq-18.5$
fm and a sizable effective range $r_{{\rm nn}}\simeq2.7$ fm may exist
in the inner crust of neutron stars \cite{Schwenk2005,Gezerlis2010,Wyk2018}.
Certainly, it is of great interest to use cold atoms to simulate and
understand any non-universal properties associated with \emph{realistic}
short-range interaction potential, particularly in thermodynamics.

In this work, we aim to establish a genuine and elegant theory to describe strongly
paired fermions with \emph{arbitrary} short-range interaction $V(r)$.
The conventional way to handle this problem is technically difficult.
For example, within mean-field theory one needs to solve the gap equation,
\begin{eqnarray}
\Delta_{{\bf k}}=-\sum_{{\bf k}^{\prime}}V_{{\bf k}{\bf k}^{\prime}}\frac{\Delta_{{\bf k}^{\prime}}}{2E_{{\bf k}^{\prime}}},
\end{eqnarray}
where $\Delta_{{\bf k}}$ is the gap function, $V_{{\bf k}{\bf k}^{\prime}}$
is the Fourier transform of the interaction potential, and $E_{{\bf k}}$
is the BCS-type single-particle dispersion. Even at the mean-field
level, this integral equation is not easy to solve \cite{MRE-01,MRE-02,MRE-03}.
Apart from numerically expensive quantum Monte Carlo (QMC) simulations,
going beyond mean-field seems impossible, since $V_{{\bf k}{\bf k}^{\prime}}$
is generally not separable. Here, our strategy is to develop an
effective field theory following the pioneering work by Kaplan \cite{Kaplan01},
who introduced a fictitious boson state and coupled it to fermions
via a contact interaction. In this way, the effective-range expansion
of the scattering phase shift is recovered in the two-body limit \cite{Kaplan01,Kaplan02}.
Interestingly, the boson state introduced by Kaplan is no longer fictitious
with the recent realization of magnetic-field-tuned Feshbach resonances
\cite{Chin2010}: it is a real dimer state in the closed channel.

The key advantage of our effective field theory is that it is constructed
to precisely reproduce the \emph{full} two-body scattering phase shift
of the potential $V(r)$ of interest, which is assumed to be known,
either directly from scattering data (i.e., for nucleon superfluids
\cite{Gezerlis2010}) or from model calculations (i.e., for quasi-two-dimensional
gases \cite{Petrov2001}). Thus, all the information of the interaction
potential is retained, beyond the effective-range expansion adopted
earlier \cite{Tajima01, Tajima02, Hu-2D-01,Hu-2D-02}. The contact boson-fermion
interaction then allows us to include crucial quantum fluctuations
beyond mean-field and hence to provide a \emph{reliable} description
of strongly paired fermions. 

As a simple application, we predict the effective-range dependence
of the ground-state energy near $s$-wave resonances in both
three (3D) and two dimensions (2D). Our results are in good agreement
with existing QMC calculations. To demonstrate the potential of using
cold atoms to understand the non-universal properties of strongly
paired fermions due to the effective-range effect and beyond, we propose
dark-state optical control of Feshbach resonances, which leads to
a non-trivial momentum dependence of the phase shift and consequently
non-universal thermodynamics. Our results pave the way to using cold
atoms to simulate realistic many-body Fermi systems, which exist ubiquitously
in all fields of physics. 

The paper is organized as follows.  In Sec. \ref{s2},  we construct a general effective Lagrangian
for an arbitrary short-range interaction.  In Sec. \ref{s3},  we formulate the many-body theory of
strongly paired fermionic superfluids based on the effective Lagrangian.  The effective-range dependence
of strongly paired fermions is studied in Sec. \ref{s4}, and an experimental scheme to probe the non-universal 
thermodynamics is proposed in Sec. \ref{s5}. We summarize in Sec. \ref{s6}. 

\section{Effective Lagrangian}\label{s2}

For the sake of simplicity, we assume that the short-range two-body interaction $V(r)$ is around an $s$-wave resonance, and thus
we neglect the contributions from higher partial waves. The $s$-wave
scattering amplitude ${\cal A}(E)$ can be expressed in terms of the
$s$-wave scattering phase shift $\delta(k)$, where $E=k^{2}/m$
is the scattering energy and $m$ is the mass of fermions ($\hbar=1$
hereafter). In 3D, we have 
\begin{eqnarray}
{\cal A}(E)=\frac{4\pi}{m}\frac{1}{k\cot\delta(k)-ik}.\label{SA-3D}
\end{eqnarray}
For a short-range interaction, $k\cot\delta(k)$ is an analytical
function of $E$, leading to the expansion 
\begin{eqnarray}
k\cot\delta(k)=-\frac{4\pi}{m}\sum_{n=0}^{\infty}c_{n}E^{n}.\label{ERE-3D}
\end{eqnarray}
Truncating to the first two terms, we obtain the so-called effective
range expansion, with $c_{0}=m/(4\pi a)$ and $c_{1}=-m^{2}r_{{\rm e}}/(8\pi)$,
where $a$ and $r_{{\rm e}}$ are the scattering length and effective
range, respectively.

A low-energy effective Lagrangian including only fermion fields $\psi_{\sigma}^{\phantom{\dag}}(x)$
($\sigma=\uparrow,\downarrow$) can be constructed order by order according to the expansion (\ref{ERE-3D}) \cite{Kaplan03}, where $x=(t,{\bf r})$ with $t$ being the time and ${\bf r}$ the 
spatial coordinates.
Here we aim to construct an effective Lagrangian that recovers exactly the scattering phase shift $\delta(k)$. Following
Kaplan \cite{Kaplan01,Kaplan02}, we introduce a boson field $\phi(x)$ that couples to the fermions. Because of the Galilean invariance, 
the general effective Lagrangian takes the form 
\begin{equation}
{\cal L}_{\textrm{eff}}=\sum_{\sigma=\uparrow,\downarrow}\psi_{\sigma}^{\dagger}\hat{K}_{{\rm F}}\psi_{\sigma}^{\phantom{\dag}}+\phi^{\dagger}{\cal F}(\hat{K}_{{\rm B}})\phi-\left(\phi^{\dagger}\psi_{\downarrow}\psi_{\uparrow}+\textrm{h.c.}\right),\label{EFT}
\end{equation}
where $\hat{K}_{{\rm F}}=i\partial_{t}+\mu+\nabla^{2}/(2m)$ and $\hat{K}_{{\rm B}}=i\partial_{t}+2\mu+\nabla^{2}/(4m)$
are the Galilean invariant kinetic operators for fermion and boson,
respectively, with $\mu$ being the chemical potential of fermions.
The crucial point of our construction is that the boson field has
an unconventional kinetic term, represented by the function ${\cal F}(\hat{K}_{{\rm B}})$,
which can be designed to recover precisely the phase shift $\delta(k)$.

To see this, let us recall that the scattering amplitude ${\cal A}(E)$
is given by the ladder summation with an interaction vertex ${\cal F}^{-1}(E)$,
which gives 
\begin{eqnarray}
{\cal A}(E)&=&-{\cal F}^{-1}(E)\sum_{n=0}^\infty\left[{\cal F}^{-1}(E){\cal B}(E)\right]^n\nonumber\\
&=&\frac{1}{{\cal B}(E)-{\cal F}(E)}.
\end{eqnarray}
Here, the two-particle bubble diagram reads 
\begin{eqnarray}
{\cal B}(E)=\sum_{{\bf p}}\frac{1}{E+i\epsilon-2\varepsilon_{{\bf p}}},
\end{eqnarray}
with $\varepsilon_{{\bf p}}={\bf p}^{2}/(2m)$ and $\sum_{\bf p}=\int d^3{\bf p}/(2\pi)^3$ for 3D. The ultraviolet (UV)
divergence in ${\cal B}$ can be regularized via a hard cutoff $\Lambda$
for $|{\bf p}|$, leading to
\begin{eqnarray}
{\cal B}(E)=-\frac{m}{4\pi}ik+{\cal D}(\Lambda),
\end{eqnarray}
where the divergent part is given by 
\begin{eqnarray}
\mathcal{D}(\Lambda)=\frac{m\Lambda}{2\pi^{2}}=\sum_{{\bf p}}\frac{1}{2\varepsilon_{{\bf p}}}.
\end{eqnarray}
Thus, the scattering amplitude takes the same form of Eq. (\ref{SA-3D}),
with the phase shift given by 
\begin{eqnarray}
k\cot\delta(k)=-\frac{4\pi}{m}{\cal F}_{{\rm R}}(E).
\end{eqnarray}
Here ${\cal F}_{{\rm R}}(E)={\cal F}(E)+{\cal D}(\Lambda)$ is the
renormalized version of the ${\cal F}$-function. It is readily seen
that the expansion Eq. (\ref{ERE-3D}) allows us to determine the
${\cal F}$-function as a polynomial in $\hat{K}_{{\rm B}}$, i.e.,
\begin{eqnarray}
{\cal F}_{{\rm R}}(\hat{K}_{{\rm B}})=\sum_{n=0}^{\infty}c_{n}(\hat{K}_{{\rm B}})^{n}.
\end{eqnarray}
Since all terms in Eq. (\ref{ERE-3D}) are included, our construction
of the effective Lagrangian Eq. (\ref{EFT}) is valid beyond the radius
of convergence of the effective-range expansion. 

The effective Lagrangian (\ref{EFT}) also applies to 2D systems, where $\nabla^{2}=\partial_x^2+\partial_y^2$. In 2D, the the $s$-wave scattering amplitude is given by \cite{scatter2D}
\begin{eqnarray}
{\cal A}(E)=\frac{4\pi}{m}\frac{1}{\pi\cot\delta(k)-i\pi}.
\end{eqnarray}
For a short-range interaction, we have
\begin{eqnarray}
\pi\cot\delta(k)=\ln\left(\frac{E}{\varepsilon_{\rm 2D}}\right)-\frac{4\pi}{m}\sum_{n=1}^\infty c_n E^n,
\end{eqnarray}
where $\varepsilon_{\rm 2D}=1/(ma_{\rm 2D}^2)$, with $a_{\rm 2D}$ being the 2D scattering length. The 2D effective range can be defined as $R_{\rm 2D}=-4\pi c_1/m^2$, which has units of length$^{2}$. Even though the leading term is nonanalytical in $E$, it is purely from the two-particle bubble diagram ${\cal B}(E)$. Using the same cutoff regularization we obtain
\begin{eqnarray}
{\cal B}(E)=-\frac{m}{4\pi}\left[\ln\left(\frac{\Lambda^2}{mE}\right)+i\pi\right],
\end{eqnarray}
Direct ladder summation shows that 
\begin{eqnarray}
\pi\cot\delta(k)=\ln\left(\frac{E}{\varepsilon_0}\right)-\frac{4\pi}{m}{\cal F}_{\rm R}(E),
\end{eqnarray}
where the the renormalized ${\cal F}$-function reads ${\cal F}_{\rm R}(E)={\cal F}(E)+{\cal D}(\Lambda)$, 
with the counter term 
\begin{eqnarray}
{\cal D}(\Lambda)=\frac{m}{4\pi}\ln\left(\frac{\Lambda^2}{m\varepsilon_0}\right)=\sum_{\bf p}\frac{1}{2\varepsilon_{\bf p}+\varepsilon_0}.
\end{eqnarray}
The energy scale $\varepsilon_0$ can be chosen arbitrarily, and we set
$\varepsilon_0=\varepsilon_{\rm 2D}$ for convenience. Thus in 2D, the ${\cal F}$-function is given by 
\begin{eqnarray}
{\cal F}_{{\rm R}}(\hat{K}_{{\rm B}})=\sum_{n=1}^{\infty}c_{n}(\hat{K}_{{\rm B}})^{n}.
\end{eqnarray}

\section{Many-Body Theory}\label{s3}

We consider a Fermi gas with a short-range two-body interaction $V(r)$. Now we can solve the many-body problem based on the effective Lagrangian
(\ref{EFT}), instead of directly using the interaction potential $V(r)$.  In the imaginary time path integral formalism, the partition function ${\cal Z}$ is given by
\begin{eqnarray}
{\cal Z}=\int[d\psi][d\psi^{\dagger}][d\phi][d\phi^{\dagger}]\exp\left[\int dx{\cal L}_{\textrm{eff}}\right],
\end{eqnarray}
where $x=(\tau,{\bf r})$ and $\int dx=\int_{0}^{\beta}d\tau\int d{\bf r}$
after the replacement $t\rightarrow-i\tau$, with $\tau$ being the
imaginary time and $\beta=1/(k_{B}T)$ the inverse temperature. The
fermions can be directly integrated out and we obtain 
\begin{eqnarray}
{\cal Z}=\int[d\phi][d\phi^{\dagger}]\exp{\Big\{ -{\cal S}_{{\rm eff}}[\phi,\phi^{\dagger}]\Big\} },
\end{eqnarray}
where the bosonic effective action reads
\begin{eqnarray}
{\cal S}_{{\rm eff}}=-\int dx\ \phi^{\dagger}{\cal F}(\hat{K}_{{\rm B}})\phi^{\phantom{\dag}}-\ {\rm Tr}\ln M_{\rm F}[\phi,\phi^{\dagger}],
\end{eqnarray}
with the fermion matrix
\begin{eqnarray}
M_{\rm F}[\phi,\phi^{\dagger}]=\left(\begin{array}{cc}
\hat{K}_{{\rm F}} & \phi\\
\phi^{\dagger} & -\hat{K}_{{\rm F}}^{*}
\end{array}\right).
\end{eqnarray}
The partition function can be alternatively expressed as
\begin{eqnarray}
{\cal Z}=\int[d\phi][d\phi^{\dagger}]\det M_{\rm F}\exp{\left[\int dx\ \phi^{\dagger}{\cal F}(\hat{K}_{{\rm B}})\phi^{\phantom{\dag}}\right]},
\end{eqnarray}
Thus the many-body problem can be simulated using the Lattice Monte Carlo method since the fermion determinant $\det M_{\rm F}$ is positive, which has been applied to the zero-range interaction case \cite{Lattice01}. 

In this work, we aim to develop an analytical theory. One advantage of the effective Lagrangian (\ref{EFT}) is that the
saddle point or classical part of the boson field $\phi$, directly
serves as the superfluid order parameter. The mean-field theory amounts
to searching for the static and uniform saddle-point solution $\phi(x)=\Delta$
that minimizes the effective action ${\cal S}_{{\rm eff}}$. In 3D and at zero
temperature, the mean-field contribution to the grand potential $\Omega=-(T/V)\ln\mathcal{Z}$
can be evaluated as 
\begin{eqnarray}
\Omega_{{\rm MF}}=\sum_{{\bf k}}\left(\xi_{{\bf k}}-E_{{\bf k}}+\frac{|\Delta|^{2}}{2\varepsilon_{{\bf k}}}\right)-|\Delta|^{2}{\cal F}_{{\rm R}}(2\mu).
\end{eqnarray}
Here $\xi_{{\bf k}}=\varepsilon_{{\bf k}}-\mu$ and $E_{{\bf k}}=(\xi_{{\bf k}}^{2}+|\Delta|^{2})^{1/2}$.
Without loss of generality, we can set $\Delta$ to be real and positive.
At $T=0$, the gap equation determining $\Delta(\mu)$ is given by
\begin{equation}
\sum_{{\bf k}}\left(\frac{1}{2E_{{\bf k}}}-\frac{1}{2\varepsilon_{{\bf k}}}\right)=-{\cal F}_{{\rm R}}(2\mu).\label{GAPE}
\end{equation}
In the strong attraction limit, the system forms a BEC of tightly
bound dimers and we have $\Delta\ll|\mu|$. The gap equation thus
reduces to a two-body equation determining the negative-energy pole
of the scattering amplitude, 
\begin{equation}
{\cal A}^{-1}(2\mu=-\varepsilon_{{\rm B}})=0,
\end{equation}
where $\varepsilon_{{\rm B}}$ is precisely the binding energy of
the dimer state determined by solving the Schroedinger equation with the interaction potential $V(r)$. In the weak attraction limit, where $\mu\simeq\varepsilon_{{\rm F}}$,
with $\varepsilon_{{\rm F}}=k_{{\rm F}}^{2}/(2m)$ being the Fermi
energy, the gap equation provides a reasonable effective-range dependence
of the pairing gap. By approximating the ${\cal F}$-function as ${\cal F}_{{\rm R}}(E)\simeq c_{0}+c_{1}E$,
the pairing gap reads 
\begin{equation}
\Delta\simeq\frac{8}{e^2}\varepsilon_{{\rm F}}\exp\left(\frac{\pi}{2k_{{\rm F}}a}-\frac{\pi}{4} k_{{\rm F}}r_{{\rm e}}\right),
\end{equation}
indicating that a positive (negative) effective range suppresses (enhances)
the pairing gap.

The mean-field theory is only qualitatively correct for strongly paired
fermions. To have a more quantitative description, we consider quantum
fluctuations around the saddle point by writing $\phi(x)=\Delta+\varphi(x)$.
The effective action can be expressed as 
\begin{equation}
{\cal S}_{{\rm eff}}=\beta V\Omega_{{\rm MF}}+{\cal S}_{{\rm FL}}\left[\varphi,\varphi^{\dagger}\right]
\end{equation}
and the partition function becomes 
\begin{equation}
{\cal Z}=e^{-\beta V\Omega_{{\rm MF}}}\int\left[d\varphi\right]\left[d\varphi^{\dagger}\right]e^{-{\cal S}_{{\rm FL}}}.
\end{equation}
An exact analytical treatment of the fluctuation contribution ${\cal S}_{{\rm FL}}$
is impossible. Here we consider only the Gaussian fluctuations,
i.e., the contributions that are quadratic in $\varphi(x)$ and $\varphi^{\dagger}(x)$,
corresponding to the contributions from collective modes. In the momentum
space, this Gaussian fluctuation contribution, ${\cal S}_{{\rm GF}}$,
can be expressed as
\begin{equation}
{\cal S}_{{\rm GF}}=\frac{1}{2}\sum_{Q}\Phi^{\dagger}(Q){\bf M}(Q)\Phi(Q),
\end{equation}
where $\Phi(Q)=[\varphi(Q),\varphi^{\dagger}(-Q)]^{{\rm T}}$. The
inverse Green's function of collective bosonic modes, ${\bf M}(Q)$,
is a $2\times2$ matrix, with elements satisfying the relations ${\bf M}_{11}(Q)={\bf M}_{22}(-Q)$
and ${\bf M}_{12}(Q)={\bf M}_{21}(Q)$. Here and in the following,
we use the notation $Q=(iq_{l},{\bf q})$, with $q_{l}=2\pi lT$ ($l\in\mathbb{Z}$)
being the bosonic Matsubara frequency. At $T=0$, the elements of
${\bf M}$ can be explicitly evaluated as 
\begin{eqnarray}
 &  & {\bf M}_{11}(Q)=\sum_{{\bf k}}\left(\frac{u_{+}^{2}u_{-}^{2}}{Y_{-}}-\frac{\upsilon_{+}^{2}\upsilon_{-}^{2}}{Y_{+}}+\frac{1}{2\varepsilon_{{\bf k}}}\right)-{\cal F}_{{\rm R}}\left(Z\right),\nonumber \\
 &  & {\bf M}_{12}(Q)=\sum_{{\bf k}}u_{+}\upsilon_{+}u_{-}\upsilon_{-}\left(\frac{1}{Y_{+}}-\frac{1}{Y_{-}}\right),
\end{eqnarray}
where $Z=iq_{l}+2\mu-{\bf q}^{2}/(4m)$, $Y_{\pm}=iq_{l}\pm(E_{+}+E_{-})$,
and the BCS distribution functions are defined as $u_{{\bf k}}^{2}=1-v_{{\bf k}}^{2}=(1+\xi_{{\bf k}}/E_{{\bf k}})/2$.
The plus and minus signs denote the momenta ${\bf k}+{\bf q}/2$ and
${\bf k}-{\bf q}/2$, respectively.

Within the Gaussian pair fluctuation (GPF) approximation \cite{TH03,TH04,TH05,TH06},
i.e., ${\cal S}_{{\rm FL}}\simeq{\cal S}_{{\rm GF}}$, the path integral
over the fluctuations can be carried out, and the grand potential is
given by $\Omega\simeq\Omega_{{\rm MF}}+\Omega_{{\rm GF}}$, where
the contribution from Gaussian fluctuations reads 
\begin{eqnarray}
\Omega_{{\rm GF}}=-\sum_{{\bf q}}\int_{-\infty}^{\infty}\frac{d\omega}{\pi}\frac{1}{e^{\beta\omega}-1}\delta_{{\rm M}}(\omega,{\bf q}),\label{GF}
\end{eqnarray}
and the phase shift $\delta_{{\rm M}}(\omega,{\bf q})=-{\rm Im}{\cal W}(\omega+i\epsilon,{\bf q})$,
with the ${\cal W}$-function given by 
\begin{eqnarray}
{\cal W}(Q)=\ln{\bf M}_{11}(Q)+\frac{1}{2}\ln\left[1-\frac{{\bf M}_{12}^{2}(Q)}{{\bf M}_{11}(Q){\bf M}_{11}(-Q)}\right].
\end{eqnarray}
The grand potential $\Omega(\mu)$ in the GPF theory can be determined,
by solving $\Delta(\mu)$ from the gap equation (\ref{GAPE}).
The density equation of state is then calculated using $n(\mu)=-\partial\Omega(\mu)/\partial\mu$.

It is worth noting that as the full scattering phase-shift is reproduced
by our theory in the two-body limit, we recover correctly the virial expansion
of the equation of state at high temperature, i.e., the Beth-Uhlenbeck
formalism for the second-order virial coefficient can be derived. At high temperature, the system is a normal gas with large but negative
$\mu$. The equation of state can be expanded in powers of the fugacity
$z=e^{\beta\mu}\ll1$ \cite{Liu2013}. In 3D we have 
\begin{equation}
{\Omega}=-\frac{2}{\beta\lambda_{T}^{3}}(z+b_{2}z^{2}+b_{3}z^{3}+\cdots),
\end{equation}
where $\lambda_{T}=\sqrt{2\pi\beta/m}$ is the thermal wavelength.
Our theory recovers the correct virial equation of state up to the
order $O(z^{2})$. To see this, we write $b_{2}=b_{2}^{(1)}+b_{2}^{(2)}$,
where the one-body contribution $b_{2}^{(1)}=-2^{-5/2}$. To find
the two-body contribution $b_{2}^{(2)}$, it is sufficient to use
Eq. (\ref{GF}), with the phase shift $\delta_{{\rm M}}(\omega,{\bf q})$
replaced by the two-body one $\delta_{{\rm 2B}}(\omega,{\bf q})$.
Using a new variable $E=\omega+2\mu-{\bf q}^{2}/(4m)$, we obtain
\begin{eqnarray}
\Omega_{2}^{(2)}=-z^{2}\sum_{{\bf q}}\int_{-\infty}^{\infty}\frac{dE}{\pi}e^{-\beta\left(E+\frac{{\bf q}^{2}}{4m}\right)}\delta_{{\rm 2B}}(E),
\end{eqnarray}
where $\delta_{{\rm 2B}}(E)=-{\rm Im}\ln[{\cal A}^{-1}(E)]$. Using
Eq. (2), we recover the elegant Beth-Uhlenbeck formalism,
\begin{eqnarray}
\frac{b_{2}^{(2)}}{\sqrt{2}}=e^{-\beta\varepsilon_{{\rm B}}}+\int_{0}^{\infty}\frac{dk}{\pi}e^{-\beta\frac{k^{2}}{m}}\frac{d\delta(k)}{dk},
\end{eqnarray}
where $\varepsilon_{{\rm B}}$ is the exact binding energy and $\delta(k)$ is the exact phase shift. 

The many-body theory for 2D is quite similar to the 3D case. Due to the energy scale $\varepsilon_{0}=\varepsilon_{{\rm 2D}}$,
the mean-field thermodynamics and gap equation are modified to,
\begin{equation}
\Omega_{{\rm MF}}=\sum_{{\bf k}}\left(\xi_{{\bf k}}-E_{{\bf k}}+\frac{\Delta{}^{2}}{2\varepsilon_{{\bf k}}+\varepsilon_{\textrm{2D}}}\right)-\Delta{}^{2}{\cal F}_{{\rm R}}(2\mu),
\end{equation}
and
\begin{equation}
\sum_{{\bf k}}\left(\frac{1}{2E_{{\bf k}}}-\frac{1}{2\varepsilon_{{\bf k}}+\varepsilon_{\textrm{2D}}}\right)=-{\cal F}_{{\rm R}}(2\mu),
\end{equation}
respectively. Moreover, for the Gaussian fluctuations, the matrix
element $\mathbf{M}_{11}(Q)$ takes the form,
\begin{equation}
{\bf M}_{11}=\sum_{{\bf k}}\left(\frac{u_{+}^{2}u_{-}^{2}}{Y_{-}}-\frac{\upsilon_{+}^{2}\upsilon_{-}^{2}}{Y_{+}}+\frac{1}{2\varepsilon_{{\bf k}}+\varepsilon_{\textrm{2D}}}\right)-{\cal F}_{{\rm R}}\left(Z\right).
\end{equation}
All the notations, i.e., $u_{\pm}^{2}$, $v_{\pm}^{2}$ and $Y_{\pm}$,
are the same as in the 3D case. 

The consistency and validity of the GPF theory has been studied for both 3D~\cite{TH03,TH04} and 2D~\cite{TH05,TH06} cases with zero-range interactions. The truncation of the pair fluctuations at the Gaussian level
provides a quantitatively good description of the BCS-BEC crossover at $T=0$, since the most important fluctuation contribution, the Goldstone mode fluctuation, is taken into account properly. 
The missing Fermi liquid correction in the BCS limit of the mean-field theory is naturally recovered by the GPF contribution~\cite{TH04}. In the BEC limit, the GPF contribution is significant to give a quantitatively good
 boson-boson interaction~\cite{TH03,TH04}. In particular, in 2D, the boson-boson interaction is missing in the mean-field theory, leading to a qualitatively incorrect equation of state in the BCS-BEC crossover. The correct boson-boson interaction is naturally recovered by the GPF contribution, leading to a correct equation of state~\cite{TH05}. In this work, the GPF theory has been generalized to finite-range interactions. We will see that the GPF theory also provides a better description of the effective-range dependence than the mean-field theory.

\section{Effective-Range Dependence}\label{s4}

\begin{figure}[t]
\centering{}\includegraphics[width=9cm]{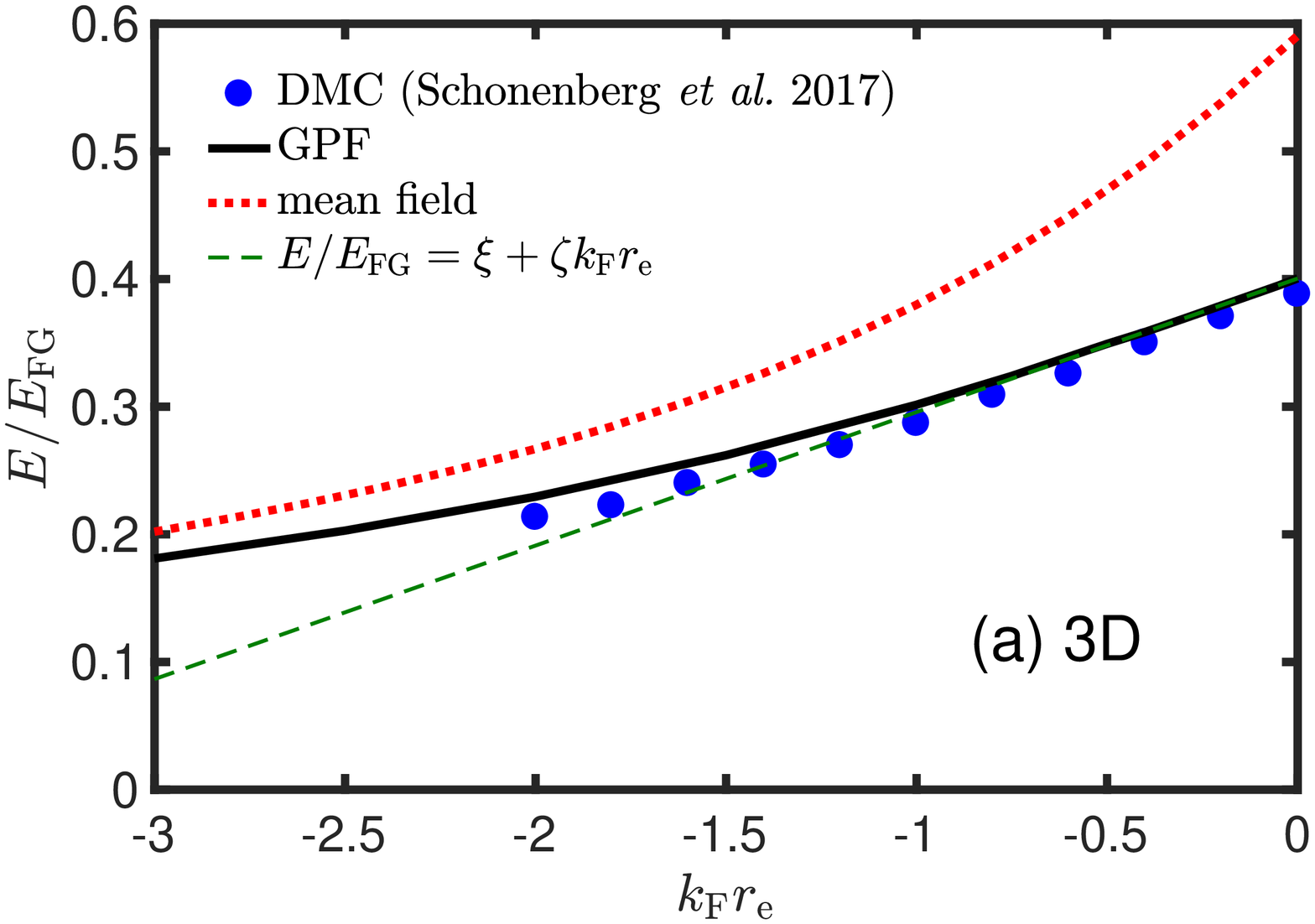}
 \includegraphics[width=9cm]{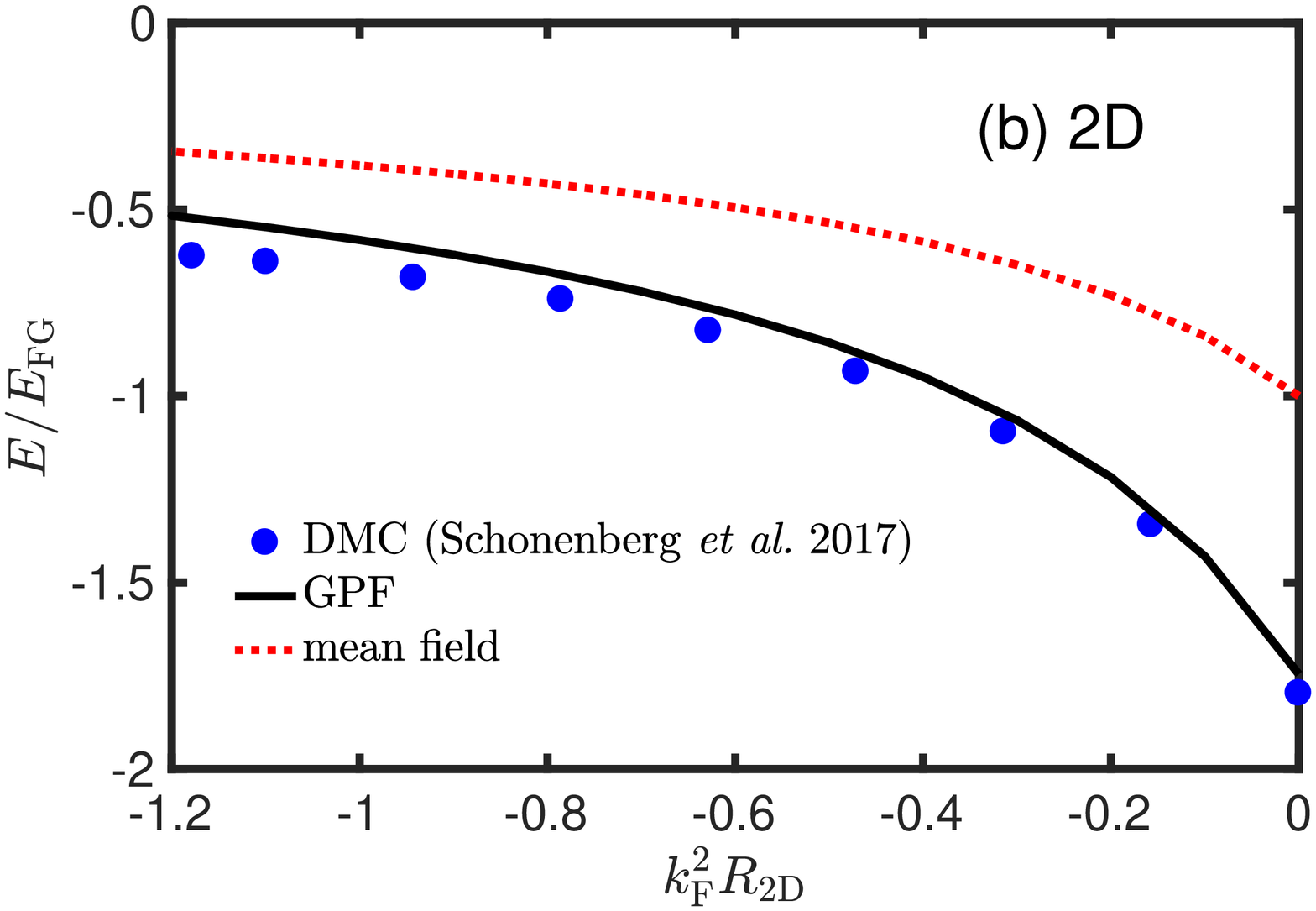}
\caption{Effective-range dependence of the ground-state energy of 3D Fermi
gases at resonance (a) and of 2D Fermi gases at vanishing mean-field
chemical potential (b). The GPF results (solid lines) are compared
with the mean-field predictions (dotted lines) and the diffusion QMC
data (solid circles) in 3D \cite{Conduit3D} and 2D \cite{Conduit2D}.
The energy is in units of the ground-state energy $E_{\ensuremath{\textrm{FG}}}$
of a non-interacting Fermi gas.\label{fig1}}
\end{figure}

We now consider the zero-temperature
equation of state of a Fermi gas with fixed density $n=k_{{\rm F}}^{3}/(3\pi^{2})$
in 3D. We focus on the effective-range dependence of the ground-state
energy at resonance, where the ${\cal F}$-function is approximated
as 
\begin{equation}
{\cal F}_{{\rm R}}(E)\simeq c_{0}+c_{1}E.
\end{equation}
While the theory in Sec. \ref{s3} applies for both positive and negative effective ranges, here we focus on a negative effective range 
since it is relevant to cold atom systems and the computational cost is small (see the Appendix).
For a negative effective range, $r_{{\rm e}}<0$, this truncation is equivalent to the two-channel
model description of the Feshbach resonance \cite{BCSBEC4}. At large
negative effective range, the model can be treated perturbatively
according to a small parameter $(k_{{\rm F}}r_{{\rm e}})^{-1}$ \cite{BCSBEC4}.
The mean-field theory provides an accurate description for $k_{{\rm F}}r_{{\rm e}}\rightarrow-\infty$.
As shown in Fig. \ref{fig1}(a), we find that the mean-field and the
GPF results converge at large $k_{{\rm F}}|r_{{\rm e}}|$, as anticipated.
At small and moderate effective range, the GPF result agrees well
with the QMC data \cite{Conduit3D}. For small $k_{{\rm F}}r_{{\rm e}}$,
the ground-state energy at resonance can be expressed as 
\begin{eqnarray}
\frac{E}{E_{{\rm FG}}}=\xi+\zeta k_{{\rm F}}r_{{\rm e}}+O\left[\left(k_{{\rm F}}r_{{\rm e}}\right)^{2}\right],
\end{eqnarray}
where the Bertsch parameter reads $\xi=0.591$ in the mean-field theory
and $\xi=0.401$ in the GPF theory. The GPF result agrees well with
the latest experimental \cite{Thomas2009,Nascimbene2010,Navon2010,Horikoshi2010,Ku2012}
and QMC \cite{EOSmc1,EOSmc2,EOSmc3,EOSmc4,Conduit3D} results, which
lie in the range $0.36-0.42$. The coefficient $\zeta$ can also be
determined. It reads $\zeta=0.273$ in the mean-field theory and $\zeta=0.105$
in the GPF theory. We note that our GPF result $\zeta=0.105$ is in
good agreement with the result $\zeta=0.087(1)$ \cite{Conduit3D}
or $\zeta=0.127(4)$ \cite{Forbes2012} from the diffusion QMC and
$\zeta=0.11(3)$ \cite{EOSmc4} from the auxiliary-field QMC.

In the BEC limit, the effective interaction between two composite dimers may be deduced.
As $\mu\rightarrow-\infty$ and $\Delta\ll|\mu|$, we expand the gap
equation (10) in powers of $\Delta/|\mu|$ and obtain,
\begin{equation}
{\cal F}_{{\rm R}}(2\mu)=\frac{m}{4\pi}\sqrt{2m|\mu|}\left(1+\frac{\Delta^{2}}{16|\mu|^{2}}\right).
\end{equation}
The solution can be expressed as $2\mu=-\varepsilon_{{\rm B}}+\mu_{{\rm B}}$,
with $\mu_{{\rm B}}\ll\varepsilon_{{\rm B}}$ being the dimer chemical
potential. We find 
\begin{equation}
\mu_{{\rm B}}=\frac{\Delta^{2}}{2\varepsilon_{{\rm B}}{\cal C}},
\end{equation}
where ${\cal C}=1+8\pi\sqrt{m\varepsilon_{{\rm B}}}{\cal F}_{{\rm R}}^{\prime}(-\varepsilon_{{\rm B}})/m^{2}$,
with ${\cal F}_{{\rm R}}^{\prime}(x)=\partial{\cal F}_{{\rm R}}(x)/\partial x$.
Meanwhile, the number equation becomes 
\begin{equation}
n=\frac{(1+\alpha){\cal C}m^{2}\Delta^{2}}{4\pi\sqrt{m\varepsilon_{{\rm B}}}},
\end{equation}
where $\alpha\sim O(1)$ comes from the Gaussian-fluctuation contribution
and depends on the details of the ${\cal F}$-function. Thus, we recover
the Bogoliubov equation of state for weakly interacting bosons, $\mu_{{\rm B}}=4\pi a_{{\rm dd}}n_{{\rm B}}/m_{{\rm B}}$,
where $m_{{\rm B}}=2m$ is the mass of the dimers and $n_{{\rm B}}=n/2$
is the density of the dimers. The dimer-dimer scattering length $a_{{\rm dd}}$
is then given by 
\begin{equation}
a_{{\rm dd}}\left(r_{{\rm e}}\right)=\frac{1}{\left(1+\alpha\right){\cal C}^{2}}\frac{2}{\sqrt{m\varepsilon_{{\rm B}}}}.
\end{equation}
Note that the above discussions are valid only for the case ${\cal C}>0$, i.e., for a repulsive dimer-dimer interaction. 

For zero-range interaction, we have $a_{{\rm dd}}(0)=2a$ from the
mean-field theory and $a_{{\rm dd}}(0)\simeq0.57a$ from the GPF theory. The GPF result is close to the exact result $a_{{\rm dd}}(0)=0.6a$ from
solving the four-body problem \cite{Petrov2004}. 
Considering only the effective range effect, i.e., ${\cal F}_{{\rm R}}(E)\simeq c_{0}+c_{1}E$,
and neglecting the weak dependence of $\alpha$ on $r_{\textrm{e}}$, we
obtain 
\begin{equation}
\frac{a_{{\rm dd}}(r_{{\rm e}})}{a_{{\rm dd}}(0)}=\frac{1+\sqrt{1-2r_{{\rm e}}/a}}{2(1-2r_{{\rm e}}/a)}.
\end{equation}
For a large negative effective range, $|r_{{\rm e}}|\gg a$, we have
$a_{{\rm dd}}\simeq a^{2}/(2|r_{{\rm e}}|)\ll a$. For a positive effective
range, $a_{{\rm dd}}$ is enhanced. The divergence at $r_{{\rm e}}=a/2$
is artificial due to our simple truncation to effective range and
is likely cured by the inclusion of the shape term $O(E^{2})$.  We also observe that the quantity ${\cal C}$ vanishes at $r_{{\rm e}}=a/2$, indicating
that the dimer-dimer interaction turns to be attractive for a larger effective range. The qualitative change around $r_{{\rm e}}\sim a$
indicates that the present analysis fails and the ground state
in this regime remains to be explored. Recent few-body calculation
shows that two dimers may form a cluster state for $r_{{\rm e}}>0.46a$
\cite{Yin2019}.

We also calculated the ground-state energy of a 2D Fermi gas with
fixed density $n=k_{{\rm F}}^{2}/(2\pi)$ as a function of the effective-range
parameter $k_{{\rm F}}^{2}R_{{\rm 2D}}$. 
In Fig. \ref{fig1}(b), we show the energy in the strongly interacting
regime where the 2D scattering length $a_{{\rm 2D}}$ is determined
by requiring $\mu=0$ within mean-field theory. If the effective-range parameter
is not large, our GPF result shows an excellent agreement with the
QMC result \cite{Conduit2D}. However, both in 2D and 3D, our GPF
predictions with the simple truncation ${\cal F}_{{\rm R}}(E)\simeq c_{0}+c_{1}E$
notably deviate from the QMC results at high density or a large effective
range, indicating that the higher-order contributions beyond the effective
range expansion may become important. These corrections could depend
sensitively on the model potentials used in QMC simulations.

\section{Probing Non-universal Thermodynamics}\label{s5}

In cold atom experiments, the $s$-wave scattering length $a$ is tuned by magnetic
field \cite{Timmermans1999,Chin2010} and the effective range is
\begin{equation}
r_{0}=-\frac{2}{ma_{{\rm bg}}\gamma B_{\Delta}},
\end{equation}
where $a_{{\rm bg}}$ is the open-channel
background scattering length, $\gamma$ is the difference of the magnetic
moment between the open and the closed channels, and $B_{\Delta}$
is the resonance width. For broad resonances in experiments, such
as $^{6}$Li at $832$G and $^{40}$K at $202$G, the effective range
$k_{{\rm F}}r_{0}$ is negligible.

\begin{figure}[t]
\centering{}\includegraphics[width=9cm]{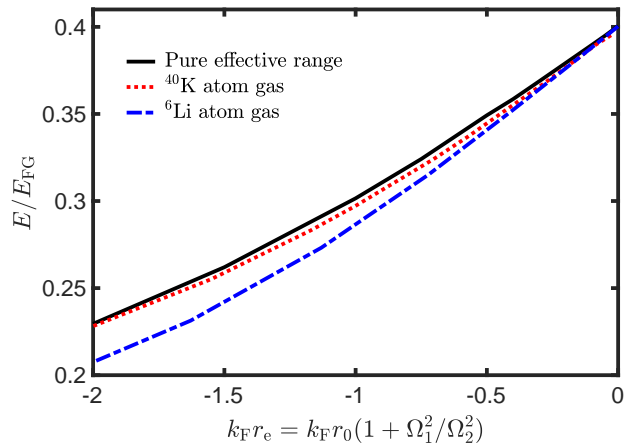} \caption{Ground-state energy of a resonantly interacting atomic Fermi gas as
a function of $k_{{\rm F}}r_{{\rm e}}$. The black solid line is the
result from the effective range expansion. The red dotted and blue
dot-dashed lines show the results for $^{40}$K and $^{6}$Li atom
gases, calculated by using the full phase shift in Eq. (\ref{ERE-DS})
under the scheme of dark-state optical control. \label{fig2}}
\end{figure}

Here we consider dark-state optical control of the Feshbach resonance
\cite{Thomas2012,Thomas2016,Thomas2018}, where two laser beams couple
the molecular state $|1\rangle$ responsible for the resonance and
another molecular state $|2\rangle$ in the closed channel to an excited
molecular state $|{\rm e}\rangle$. Near the resonance, the scattering
phase shift is modified to \cite{Thomas2012,Zhang2017,He2018} 
\begin{equation}
k\cot\delta(k)=-\frac{1}{a_{{\rm bg}}}\frac{E-\Sigma(E)}{E-\Sigma(E)+\gamma B_{\Delta}},\label{ERE-DS}
\end{equation}
where $\Sigma(E)$ is the optically induced Stark shift. In the dark-state
regime, we have $\Sigma(0)=0$ and $\Sigma^{\prime}(0)=-\Omega_{1}^{2}/\Omega_{2}^{2}$,
where $\Omega_{1}$ and $\Omega_{2}$ are the Rabi frequencies for
the transitions $|1\rangle\leftrightarrow|{\rm e}\rangle$ and $|2\rangle\leftrightarrow|{\rm e}\rangle$.
Thus the resonance does not shift but the effective range can be tuned
by changing the ratio $\Omega_{1}/\Omega_{2}$ \cite{Thomas2012}:
\begin{equation}
r_{{\rm e}}=r_{0}\left(1+\frac{\Omega_{1}^{2}}{\Omega_{2}^{2}}\right).
\end{equation}

Figure \ref{fig2} shows the ground-state energy as a function of
$k_{{\rm F}}r_{{\rm e}}$ for $^{6}$Li and $^{40}$K atom gases,
at resonances $832$ G and $202$ G, respectively. The density $n$
is chosen as $2\times10^{14}$ cm$^{-3}$, a typical value realized
in cold atom experiments. The effective range parameter without optical
control is $k_{{\rm F}}r_{0}=-4.5\times10^{-4}$ for $^{6}$Li and
$k_{{\rm F}}r_{0}=-2.7\times10^{-2}$ for $^{40}$K. We find that the
result from a $^{40}$K atom gas agrees well with the effective range
expansion. This is because the resonance for $^{40}$K has a relatively
large intrinsic effective range $r_{0}$. To reach $k_{{\rm F}}r_{{\rm e}}\sim O(1)$,
we need $\Omega_{1}/\Omega_{2}\sim10$ for $^{40}$K and a much larger
value $\Omega_{1}/\Omega_{2}\sim100$ for $^{6}$Li. As a result,
for $^{6}$Li the higher-order terms beyond the effective range expansion
of Eq. (\ref{ERE-DS}) become important. 

\section{Summary}\label{s6}

We have proposed a convenient way to describe strongly paired fermions without knowing the details of the
short-range interaction potential. Within our effective field theory,
it is easy to go beyond mean field to include quantum fluctuations
for arbitrary potential. The predicted effective-range dependence
of the ground-state energy is in good agreement with recent quantum
Monte Carlo simulations. At a large effective range, the effect beyond
the effective-range expansion may become significant. Our theory can
be readily generalized to include the contributions from higher partial
waves \cite{Yu2016,Yu2017} and extended to study bosonic systems \cite{You2017,Pan2019}. 
It is also interesting to study the case of a positive effective range, which is relevant to the equation of state at low density neutron matter 
of astrophysical interest~\cite{Schwenk2005,Gezerlis2010,Wyk2018}.

\begin{acknowledgments}
We thank Xiangyu Yin and Ren Bi for useful discussions. This research
was supported by the National Natural Science Foundation of China,
Grant Nos. 11775123 and 11890712 (L.H.), National Key Research and Development Program
of China, Grant No. 2018YFA0306503 (L.H.), and Australian Research
Council's (ARC) Discovery Program, Grant No. FT140100003 (X.-J.L),
Grant No. DP180102018 (X.-J.L), and Grant No. DP170104008 (H.H.).
\end{acknowledgments}

\begin{widetext}
\appendix

\section{Evaluating the fluctuation contribution $\Omega_{\rm GF}$ at $T=0$}

At $T=0$, there is a much easier approach to evaluate $\Omega_{\rm GF}$ for some cases proposed by Diener, Sensarma, and Randeria~\cite{TH04}. We define two functions ${\bf M}^{\rm C}_{11}$ and  ${\bf M}^{\rm C}_{22}$, which in 3D are given by
\begin{eqnarray}
{\bf M}^{\rm C}_{11}(Q)=-{\cal F}_{\rm R}(Z)+\sum_{\bf k}\left(\frac{u_+^2u_-^2}{iq_l-E_+-E_-}+\frac{1}{2\varepsilon_{\bf k}}\right)
\end{eqnarray}
and ${\bf M}^{\rm C}_{22}(Q)={\bf M}^{\rm C}_{11}(-Q)$. In 2D, we need to replace $1/(2\varepsilon_{\bf k})$ with $1/(2\varepsilon_{\bf k}+\varepsilon_{\rm 2D})$. At $T=0$,  the boson Matsubara sum is converted to a continuous integral over an imaginary frequency $i\omega$, i.e.,
\begin{equation}
 \frac{1}{\beta}\sum^{\infty}_{l=-\infty}F(iq_{l})=\int^{\infty}_{-\infty}\frac{d\omega}{2\pi}F(i\omega).
\end{equation}
Therefore, if the Matsubara sum $\sum_{l}\ln {\bf M}^{\rm C}_{11}(Q)$ vanishes even without the convergent factor $e^{iq_{l}0^{+}}$ at $T=0$, we arrive at an alternative convergent expression for 
$\Omega_{\rm GF}$~\cite{TH04}:
\begin{eqnarray}
 \Omega_{\rm GF}=\sum_{\bf q}\int^{\infty}_{0}\frac{d\omega}{2\pi}
 \ln\left[\frac{{\bf M}_{11}(\omega,{\bf q}){\bf M}_{11}(-\omega,{\bf q})-{\bf M}_{12}^2(\omega,{\bf q})}
{{\bf M}_{11}^{\rm C}(\omega,{\bf q}){\bf M}_{11}^{\rm C}(-\omega,{\bf q})}\right].
\end{eqnarray}
Here we have used the fact that the integrand is even in $\omega$. 

If we consider only the effective-range effect and approximate the ${\cal F}$-function as 
${\cal F}_{{\rm R}}(E)\simeq c_{0}+c_{1}E$, the counter function in 3D can be expressed as
\begin{eqnarray}
{\bf M}^{\rm C}_{11}(Q)=\frac{m^2r_{\rm e}}{8\pi}\left(iq_l-\frac{{\bf q}^2}{4m}\right)+\int\frac{d^3{\bf k}}{(2\pi)^3}\left(\frac{u_+^2u_-^2}{iq_l-E_+-E_-}+\frac{1}{2E_{\bf k}}\right).
\end{eqnarray}
For negative effective range ($r_{\rm e}<0$),  the function ${\bf M}_{11}^{\rm C}(z,{\bf q})$ has no zeros in the left half plane (${\rm Re}z<0$) for an arbitrary value of ${\bf q}$.  Therefore, the Matsubara sum 
$\sum_{l}\ln {\bf M}^{\rm C}_{11}(Q)$ vanishes at $T=0$ and the above trick applies. After some manipulations, we obtain
\begin{eqnarray}
\Omega_{\rm GF}=\sum_{\bf q}\int_0^\infty\frac{d\omega}{2\pi}
\ln\left[1-2\Delta^4\frac{A(\omega,{\bf q})C(\omega,{\bf q})+\omega^2B(\omega,{\bf q})D(\omega,{\bf q})
+2F^2(\omega,{\bf q})}{A^2(\omega,{\bf q})+\omega^2B^2(\omega,{\bf q})}+\Delta^8\frac{C^2(\omega,{\bf q})+\omega^2D^2(\omega,{\bf q})}
{A^2(\omega,{\bf q})+\omega^2B^2(\omega,{\bf q})}\right].\nonumber\\
\end{eqnarray}
The functions $A,B,C,D,$ and $F$ are defined as
\begin{eqnarray}\label{GPFfun}
&&A(\omega,{\bf q})=\frac{m^2r_{\rm e}}{8\pi}\left(2\mu-\frac{{\bf q}^2}{4m}\right)+
\int\frac{d^3{\bf k}}{(2\pi)^3} \left[\frac{1}{2E_{\bf k}}
-\frac{1}{4}\left(\frac{1}{E_+}+\frac{1}{E_-}\right)\frac{(E_++\xi_+)(E_-+\xi_-)}{(E_++E_-)^2+\omega^2}\right],\nonumber\\
&&B(\omega,{\bf q})=-\frac{m^2}{8\pi}r_{\rm e}+\int\frac{d^3{\bf k}}{(2\pi)^3} \frac{1}{4E_+E_-}
\frac{(E_++\xi_+)(E_-+\xi_-)}{(E_++E_-)^2+\omega^2},\nonumber\\
&&C(\omega,{\bf q})=\int\frac{d^3{\bf k}}{(2\pi)^3} \frac{1}{4}\left(\frac{1}{E_+}+\frac{1}{E_-}\right)
\frac{1}{(E_++\xi_+)(E_-+\xi_-)}\frac{1}{(E_++E_-)^2+\omega^2},\nonumber\\
&&D(\omega,{\bf q})=\int\frac{d^3{\bf k}}{(2\pi)^3} \frac{1}{4E_+E_-(E_++\xi_+)(E_-+\xi_-)}
\frac{1}{(E_++E_-)^2+\omega^2},\nonumber\\
&&F(\omega,{\bf q})=\int\frac{d^3{\bf k}}{(2\pi)^3} \frac{1}{4}\left(\frac{1}{E_+}+\frac{1}{E_-}\right)\frac{1}{(E_++E_-)^2+\omega^2}.
\end{eqnarray}
This expression for $\Omega_{\rm GF}$ leads to a rapidly convergent result.  However, for positive effective range ($r_{\rm e}>0$),  the function ${\bf M}_{11}^{\rm C}(z,{\bf q})$ has zeros in the left half plane for sufficiently large values of $|{\bf q}|$. In this case, the above trick fails and we need to use the phase shift expression (\ref{GF}). The convergence of the integral over the real frequency is rather slow and the computational cost becomes large.
Similar discussions also apply to the 2D case.

\end{widetext}


\begin{thebibliography}{10}
\bibitem{Review01} {D. Vollhardt and P. Woelfle, \emph{The Superfluid
Phases of Helium 3} (Taylor and Francis, London, 1990).} 

\bibitem{Review02} {S. Giorgini, L. P. Pitaevskii, and S. Stringari,
Rev. Mod. Phys. \textbf{80}, 1215 (2008); I. Bloch, J. Dalibard, and
W. Zwerger, Rev. Mod. Phys. \textbf{80}, 885 (2008).} 

\bibitem{Review03} {S. Gandolfi, A. Gezerlis, and J. Carlson, Annu.
Rev. Nucl. Part. Sci. \textbf{65}, 303 (2015); A. Sedrakian and J.
W. Clark, arXiv:1802.00017; G. C. Strinati, P. Pieri, G. Roepke, P.
Schuck, and M. Urban, Phys. Rep. \textbf{738}, 1 (2018).}

\bibitem{Exciton} {A. Perali, D. Neilson, and A. R. Hamilton, Phys.
Rev. Lett. \textbf{110}, 146803 (2013); P. Lopez Rios, A. Perali, R.
J. Needs, D. Neilson, Phys. Rev. Lett. \textbf{120}, 177701 (2018);
G. W. Burg, N. Prasad, K. Kim, T. Taniguchi, K. Watanabe, A. H. MacDonald,
L. F. Register, and E. Tutuc, Phys. Rev. Lett. \textbf{120}, 177702
(2018).} 

\bibitem{Eagles} {D. M. Eagles, Phys. Rev. \textbf{186}, 456 (1969).} 

\bibitem{Leggett} {A. J. Leggett, in \emph{Modern Trends in the
Theory of Condensed Matter}, \emph{Lecture Notes in Physics}, Vol.
\textbf{115} (Springer-Verlag, Berlin, 1980).} 

\bibitem{NSR} {P. Nozieres and S. Schmitt-Rink, J. Low Temp. Phys.
\textbf{59}, 195 (1985).} 

\bibitem{BCSBEC1} {C. A. R. Sa de Melo, M. Randeria, and J. R. Engelbrecht,
Phys. Rev. Lett. \textbf{71}, 3202 (1993).} 

\bibitem{BCSBEC2} {J. R. Engelbrecht, M. Randeria, and C. A. R.
Sa de Melo, Phys. Rev. B \textbf{55}, 15153 (1997).} 

\bibitem{BCSBEC3} {Q. Chen, J. Stajic, S. Tan, and K. Levin, Phys.
Rep. \textbf{412}, 1 (2005).} 

\bibitem{BCSBEC4} {V. Gurarie, and L. Radzihovsky, Ann. Phys. (N.
Y.) \textbf{322}, 2 (2007).} 

\bibitem{Regal2004} {C. A. Regal, M. Greiner, and D. S. Jin. Phys.
Rev. Lett. \textbf{92}, 040403 (2004).}

\bibitem{Zwierlein2004} {M. W. Zwierlein, C. A. Stan, C. H. Schunck,
S. M. F. Raupach, A. J. Kerman, and W. Ketterle, Phys. Rev. Lett.
\textbf{92}, 120403 (2004).}

\bibitem{Kinast2004} {J. Kinast, S. L. Hemmer, M. E. Gehm, A. Turlapov,
and J. E. Thomas, Phys. Rev. Lett. \textbf{92}, 150402 (2004).}

\bibitem{Ho2004}{T.-L. Ho, Phys. Rev. Lett. \textbf{92}, 090402
(2004).}

\bibitem{Hu2007} {H. Hu, X. -J. Liu, and P. D. Drummond, Nat. Phys.
\textbf{3}, 469 (2007).} 

\bibitem{Thomas2009} {L. Luo and J. E.Thomas, J. Low Temp. Phys.
\textbf{154}, 1 (2009).}

\bibitem{Nascimbene2010} {S. Nascimbene, N. Navon, K. J. Jiang,
F. Chevy, and C. Salomon, Nature (London) \textbf{463}, 1057 (2010).} 

\bibitem{Horikoshi2010} {M. Horikoshi, S. Nakajima, M. Ueda, and
T. Mukaiyama, Science \textbf{327}, 442 (2010).} 

\bibitem{Navon2010} {N. Navon, S. Nascimbene, F. Chevy, and C. Salomon,
Science \textbf{328}, 729 (2010).} 

\bibitem{Ku2012} {M. J. H. Ku, A. T. Sommer, L. W. Cheuk, and M.
W. Zwierlein, Science \textbf{335}, 563 (2012).}

\bibitem{TH01} {A. Perali, P. Pieri, L. Pisani, and G. C. Strinati,
Phys. Rev. Lett. \textbf{92}, 220404 (2004).} 

\bibitem{TH02} {P. Pieri, L. Pisani, and G. C. Strinati, Phys. Rev.
B \textbf{70}, 094508 (2004).} 

\bibitem{TH03} {H. Hu, X.-J. Liu, and P. D. Drummond, Europhys.
Lett. \textbf{74}, 574 (2006).} 

\bibitem{TH04} {R. B. Diener, R. Sensarma, and M. Randeria, Phys.
Rev. A \textbf{77}, 023626 (2008).} 

\bibitem{TH05} {L. He, H. Lu, G. Cao, H. Hu, and X.-J. Liu, Phys.
Rev. A \textbf{92}, 023620 (2015).} 

\bibitem{TH06} { L. Salasnich and F. Toigo, Phys. Rev. A {\bf 91}, 011604 (2015).}

\bibitem{TH07} {Y. Nishida and D. T. Son, Phys. Rev. Lett. \textbf{97},
050403 (2006).} 

\bibitem{TH08} {R. Haussmann, W. Rantner, S. Cerrito, and W. Zwerger,
Phys. Rev. \textbf{A75}, 023610 (2007).} 

\bibitem{TH09} {M. Y. Veillette, D. E. Sheehy, and L. Radzihovsky,
Phys. Rev. A \textbf{75}, 043614 (2007).} 

\bibitem{TH10} {E. Taylor, A. Griffin, N. Fukushima, and Y. Ohashi,
Phys. Rev. A \textbf{74}, 063626 (2006).} 

\bibitem{TH11} {B. C. Mulkerin, L. He, P. Dyke, C. J. Vale, X.-J. Liu, and H. Hu, Phys. Rev. A {\bf 96}, 053608 (2017).}


\bibitem{TH12} {G. Bighin and L. Salasnich, Phys. Rev. B {\bf 93}, 014519 (2016).}


\bibitem{EOSmc1} {J. Carlson, S.-Y. Chang, V. R. Pandharipande,
and K. E. Schmidt, Phys. Rev. Lett. \textbf{91}, 050401 (2003).} 

\bibitem{EOSmc2} {C. Lobo, A. Recati, S. Giorgini, and S. Stringari,
Phys. Rev. Lett. \textbf{97}, 200403 (2006).} 

\bibitem{EOSmc3} {M. M. Forbes, S. Gandolfi, and A. Gezerlis, Phys.
Rev. Lett. \textbf{106}, 235303 (2011).} 

\bibitem{EOSmc4} {J. Carlson, S. Gandolfi, K. E. Schmidt, and S.
Zhang, Phys. Rev. A \textbf{84}, 061602(R) (2011).} 

\bibitem{Schwenk2005}   {A. Schwenk and C. J. Pethick, Phys. Rev. Lett. {\bf 95}, 160401 (2005).}

\bibitem{Gezerlis2010} {A. Gezerlis and J. Carlson, Phys. Rev. C
\textbf{81}, 025803 (2010).} 

\bibitem{Wyk2018} {P. van Wyk, H. Tajima, D. Inotani, A. Ohnishi,
and Y. Ohashi, Phys. Rev. A \textbf{97}, 013601 (2018).}

\bibitem{MRE-01} {N. Andrenacci, A. Perali, P. Pieri, and G. C.
Strinati, Phys. Rev. B \textbf{60}, 12410 (1999).} 

\bibitem{MRE-02} {M. M. Parish, B. Mihaila, E. M. Timmermans, K.
B. Blagoev, and P. B. Littlewood, Phys. Rev. B \textbf{71}, 064513
(2005).} 

\bibitem{MRE-03} {L. M. Jensen, H. M. Nilsen, and G. Watanabe, Phys.
Rev. A \textbf{74}, 043608 (2006).} 

\bibitem{Kaplan01} {D. B. Kaplan, Nucl. Phys. B \textbf{494}, 471
(1997).} 

\bibitem{Kaplan02} {D. B. Kaplan and S. Sun, Phys. Rev. Lett. \textbf{107},
030601 (2011).}

\bibitem{Chin2010} {C. Chin, R. Grimm, P. Julienne, and E. Tiesinga,
Rev. Mod. Phys. \textbf{82}, 1225 (2010).}

\bibitem{Petrov2001}D. S. Petrov and G. V. Shlyapnikov, Phys. Rev.
A \textbf{64}, 012706 (2001).

\bibitem{Hu-2D-01} {H. Hu, B. C. Mulkerin, U. Toniolo, L. He, and
X.-J. Liu, Phys. Rev. Lett. \textbf{122}, 070401 (2019).} 

\bibitem{Tajima01}   {H. Tajima, Phys. Rev. A {\bf 97}, 043613 (2018).}

\bibitem{Tajima02}   {H. Tajima, J. Phys. Soc. Jpn. {\bf 88}, 093001 (2019).}

\bibitem{Hu-2D-02} {F. Wu, J. Hu, L. He, X.-J. Liu, and H. Hu, arXiv:1906.08578.}

\bibitem{Kaplan03}   {D. B. Kaplan, M. J. Savage, and M. B. Wise, Nucl. Phys. B{\bf 478}, 629 (1996).}


\bibitem{scatter2D} {S. K. Adhikari, Am. J. Phys. \textbf{54}, 362
(1986).} 

\bibitem{Liu2013} {X.-J. Liu, Phys. Rep. \textbf{524}, 37 (2013).} 

\bibitem{Lattice01}  {M. G. Endres, D. B. Kaplan, J.-W. Lee, and A. N. Nicholson, Phys. Rev. A {\bf 84}, 043644 (2011); Phys. Rev. A {\bf 87}, 023615 (2013).}

\bibitem{Conduit3D} {L. M. Schonenberg and G. J. Conduit, Phys.
Rev. A \textbf{95}, 013633 (2017).} 

\bibitem{Forbes2012} {M. M. Forbes, S. Gandolfi, and A. Gezerlis,
Phys. Rev. A \textbf{86}, 053603 (2012).} 


\bibitem{Petrov2004}     {D. S. Petrov, C. Salomon, and G. V. Shlyapnikov, Phys. Rev. Lett. {\bf 93}, 090404 (2004).}

\bibitem{Yin2019} {X. Y. Yin, H. Hu, and X.-J. Liu, Phys. Rev. Lett. {\bf 123}, 073401 (2019). } 

\bibitem{Conduit2D} {L. M. Schonenberg, P. C. Verpoort, and G. J.
Conduit, Phys. Rev. A \textbf{96}, 023619 (2017).} 


\bibitem{Timmermans1999} {E. Timmermans, T. Tomassini, M. Hussein,
and A. Kerman, Phys. Rep. \textbf{315}, 199 (1999).} 

\bibitem{Thomas2012} {H. Wu and J. E. Thomas, Phys. Rev. Lett. \textbf{108},
010401 (2012); Phys. Rev. A \textbf{86}, 063625 (2012).} 

\bibitem{Thomas2016} {A. Jagannathan, N. Arunkumar, J. A. Joseph,
and J. E. Thomas, Phys. Rev. Lett. \textbf{116}, 075301 (2016).} 

\bibitem{Thomas2018} {N. Arunkumar, A. Jagannathan, and J. E. Thomas,
Phys. Rev. Lett. \textbf{121}, 163404 (2018).} 

\bibitem{Zhang2017} {J. Jie and P. Zhang, Phys. Rev. A \textbf{95},
060701(R) (2017).} 

\bibitem{He2018} {L. He, H. Hu, and X.-J. Liu, Phys. Rev. Lett.
\textbf{120}, 045302 (2018).}

\bibitem{Yu2016}   { Z. Yu, J. H. Thywissen, and S. Zhang, Phys. Rev. Lett. {\bf 115}, 135304 (2015).}

\bibitem{Yu2017}   {P. Zhang, S. Zhang, and Z. Yu, Phys. Rev. A {\bf 95}, 043609 (2017).}

\bibitem{You2017}   {Y. Cui \emph{et al.}, Phys. Rev. Lett. {\bf 119}, 203402 (2017).}

\bibitem{Pan2019}  {X.-C. Yao \emph{et al.}, Nat. Phys. {\bf 15}, 570 (2019).}

\end{thebibliography}
\end{document}